\documentclass[aps,prl,twocolumn]{revtex4}
\usepackage{amssymb,lipsum}
\usepackage{graphicx,amsmath,color}

\newcommand{\pdag}{{\phantom{\dagger}}}
\newcommand{\bea}{\begin{eqnarray}}
\newcommand{\eea}{\end{eqnarray}}
\newcommand{\beq}{\begin{equation}}
\newcommand{\eeq}{\end{equation}}
\newcommand{\benu}{\begin{enumerate}}
\newcommand{\enu}{\end{enumerate}}

\newcommand{\om}{\omega}
\newcommand{\Om}{\Omega}
\newcommand{\ep}{\epsilon}

\newcommand{\bk}{{\bf k}}
\newcommand{\bq}{{\bf q}}

\newcommand{\br}{{\bf r}}

\begin{document}

\title{
Lattice effects on nematic quantum criticality in metals
}

\author{I. Paul$^1$ and M. Garst$^2$}

\affiliation{
$^1$Laboratoire Mat\'{e}riaux et Ph\'{e}nom\`{e}nes Quantiques, Universit\'{e} Paris Diderot-Paris 7 \& CNRS,
UMR 7162, 75205 Paris, France \\
$^2$Institut f\"{u}r Theoretische Physik, Technische Universit\"{a}t Dresden, 01062 Dresden, Germany
}

\begin{abstract}
Theoretically, it is commonly held that in metals near a nematic quantum critical point
the electronic excitations become incoherent on the entire ``hot'' Fermi surface, triggering
non Fermi liquid behavior. However, such conclusions are based on electron-only theories,
ignoring a symmetry-allowed coupling between the electronic nematic variable and a suitable crystalline lattice strain.
Here we show that including this coupling leads to entirely
different conclusions because the critical fluctuations are mostly cutoff by the non-critical lattice shear modes.
At sufficiently low temperatures the thermodynamics remain Fermi liquid type, while, depending on the Fermi surface
geometry, either the entire Fermi surface stays cold, or at most there are hot spots.
In particular, our predictions are relevant for the iron-based superconductors.
\end{abstract}

\date{\today}

\maketitle
%%%%%%%%%%%%%%%%%%%%%%%%%%%%%%%%%%%%%%%%%%%%%%%%%%%%%%%%%%%%%%%%%%%%%%%%%
%%%%%%%%%%%%%%%%%%%%%%%%%%%%%%%%%%%%%%%%%%%%%%%%%%%%%%%%%%%%%%%%%%%%%%%%%%%%%%%%%%%%%%%%%%%%%%%%%%%%
At an Ising nematic quantum critical point (QCP) in solids,
discussed often in the context of the iron-based superconductors,
cuprates, ruthanates, and quantum Hall systems,
the ground state transforms from one
having discrete rotational symmetry to another in which this symmetry is broken
(see Figure~1)~\cite{Fradkin2010,Nie2013,Achkar2016,Borzi2007,Lilly1999,Chu2010,Fernandes2014,Gallais2016}.
An ideal example is the tetragonal to orthorhombic structural transition at temperature $T_S$ in the
iron superconductors (FeSC), which is driven by electronic correlations, and where $T_S \rightarrow 0$
with doping~\cite{Chu2010,Fernandes2014,Gallais2016,Johnston2010}.
Besides the FeSC, a nematic QCP is often invoked in the context of several other correlated
metals, notably the cuprates~\cite{Fradkin2010,Nie2013,Achkar2016}. Consequently, a topic of immediate
relevance is
how the quantum fluctuations associated with this QCP affect the low temperature properties of a metal.

At present it is widely believed that the effective electron-electron interaction becomes long-ranged near the
nematic QCP~\cite{Lohneysen2007,Oganesyan2001,Metzner2003,Garst2010}.
As a result the electrons become unusually massive and short-lived,
leading to
non Fermi liquid (NFL) behavior both in thermodynamics and in single electron properties
almost everywhere on the Fermi surface.
Thus, the specific heat coefficient $\gamma \equiv -\partial^2 F/\partial T^2$, where $F(T)$ is the free energy, diverges
as $\gamma(T) \propto 1/T^{1/3}$ in space dimension $d=2$, and as $\gamma \propto \log T$ in $d=3$.
Simultaneously, almost the entire Fermi surface gets ``hot'', and is characterized by a frequency dependent self-energy
$\Sigma(i \omega_n) \propto |\omega_n|^{2/3}$ in $d=2$, and by $\Sigma(i \omega_n) \propto \omega_n \log|\omega_n|$
in $d=3$.

%====================
\begin{figure}[!!t]
\begin{center}
\includegraphics[width=0.95\linewidth,trim=0 0 0 0]{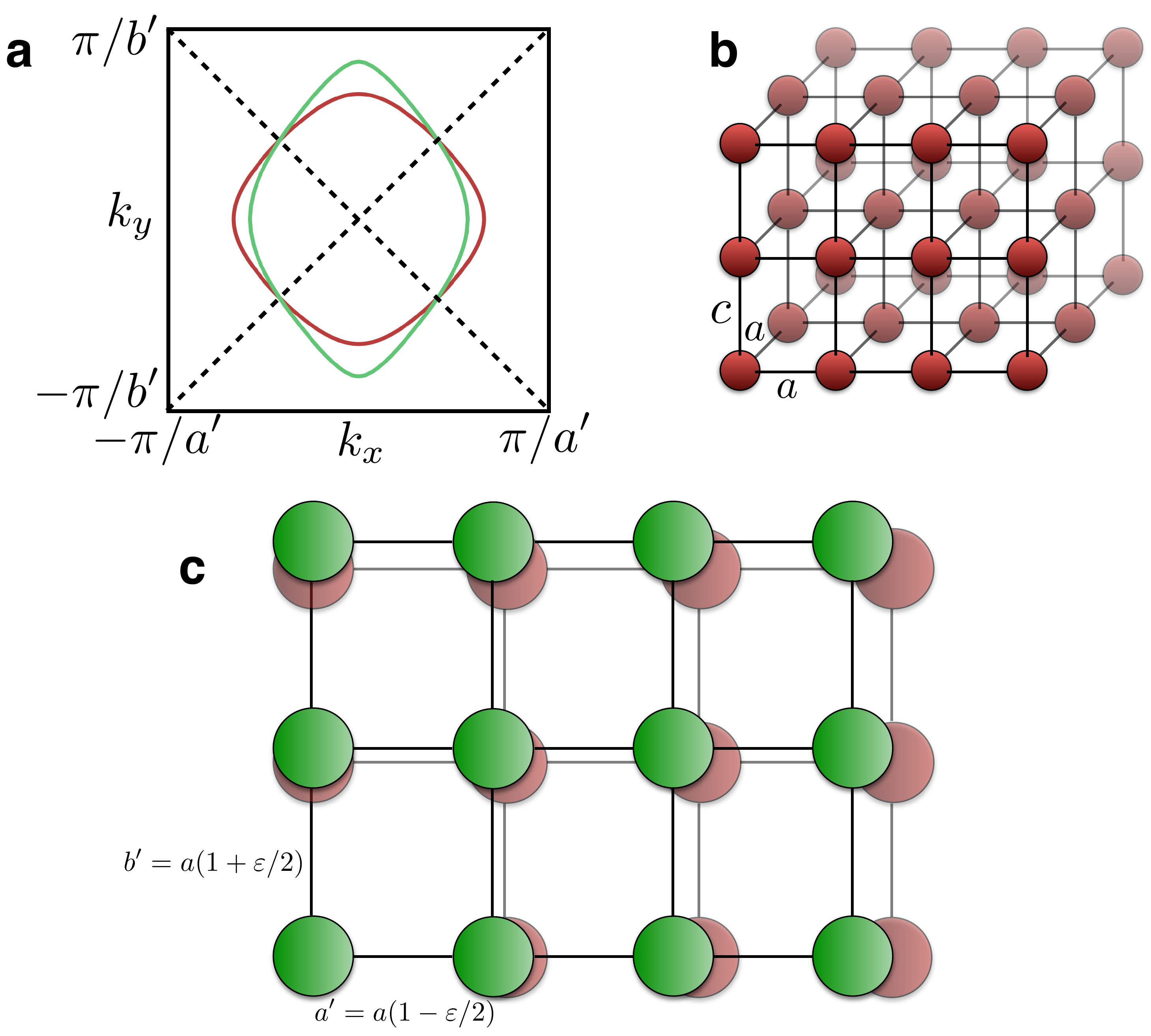}
\caption{
\textbf{Ising nematic phase transition involving $(x^2 - y^2)$ symmetry breaking.}
(a) The $C_4$ symmetric Fermi surface
(red) distorts and becomes $C_2$ symmetric (green) in the nematic phase. (b) A tetragonal
lattice with equivalent $\hat{x}$ and $\hat{y}$ directions. (c) View of its $x-y$ plane,
which distorts in the nematic phase, red and green circles being the original
and the distorted atomic positions respectively. In the nematic phase the unit cell lengths
$a^{\prime}$ and $b^{\prime}$ along the two directions become inequivalent.
$\epsilon$ is the orthorhombic strain. Even if the electron dispersion is two-dimensional, in
the presence of the lattice the third dimension is important (see text).
}
\label{fig1}
\end{center}
\end{figure}
%====================

These results are based on the simplest treatment of the typical model describing itinerant electrons interacting with the
critical nematic collective mode of the electrons themselves~\cite{Lohneysen2007}. The latter is characterized by a susceptibility
\beq
\label{eq:chi-n}
\chi_0^{-1} (\bq, i \Omega_n) = \nu_0^{-1} \left[ r + q^2 + D(\bq, i \Om_n) \right],
\eeq
where $\nu_0$ is a constant with dimension of density of states, and $\bq$ and $\Om_n$ are dimensionless momentum and
Matsubara frequency,
respectively. The dynamics of the collective mode is damped due to the excitation of particle-hole pairs close to the Fermi
surface, $D(\bq, i \Om_n) \propto |\Om_n|/q$.
At the QCP the tuning parameter vanishes, $r=0$.

More recently, a lot of work has been done to improve the theory
in $d=2$~\cite{Lee2009,Metlitski2010,Mross2010,Schattner2015,Drukier2012,Holder2015}.
However, these works do not question the belief
that the electronic properties are NFL type. In fact, it is widely accepted
that quantum criticality involving a non-modulating order parameter invariably leads to NFL physics.

Experimentally, the existence of NFL physics is well-established in the pseudogap and strange metal phases
of the cuprates~\cite{cuprate-nfl}. NFL behavior has also been reported for certain, if not all,
FeSC~\cite{nfl-fesc}. However, at present
there is no definite evidence that the NFL physics is due to a nematic QCP, since there are other possible sources
of NFL behavior, such as spin fluctuations and Mott physics.

\textbf{Nemato-elastic coupling.}
The link between nematic QCP and NFL behavior is based on an electron-only theory.
In practice, in a solid the electronic environment
is sensitive to the lattice strains, and this gives rise to a symmetry-allowed nemato-elastic coupling between the
electron-nematic variable $\phi$ and a suitable component of the strain tensor of the type
\beq
\label{eq:nem-latt-coupling}
\mathcal{H}_{\rm nem-latt} = \lambda \int d{\bf r} \phi (\bf {r}) \varepsilon(\bf {r}),
\eeq
where $\lambda$ is the coupling constant with dimension of energy.
For the sake of concreteness we assume $\phi$ to
transform as $(x^2- y^2)$ under the point group operations.
Then,
$\varepsilon({\bf r}) = \varepsilon + i \sum_{\bq \neq 0}
\left[ q_x u_x(\bq) - q_y u_y(\bq) \right] e^{i \bq \cdot \br}$
is the local orthorhombic strain, $\vec u({\bf r})$ is the atomic displacement associated with
strain fluctuation, and
the uniform macroscopic strain $\varepsilon \neq 0$ in the symmetry-broken
nematic/orthorhombic phase, see Fig.~\ref{fig1}(c).
The problem is well-posed if we assume that the undistorted
lattice is tetragonal, whose elastic energy is given by
$F_E = \int d^d r C_{ijkl} \varepsilon_{ij}(r) \varepsilon_{kl}(r)/2$, $i = (x, y, z)$,
where $C_{ijkl}$ are the bare elastic constants (for an explicit expression in the more convenient Voigt notation
used henceforth, see Supplementary Information (SI))~\cite{Landau-Lifshitz,supplemental}.

Importantly, the above coupling shifts the nematic QCP, and it occurs already at a finite value of $r$ given by
\beq
\label{eq:r0}
r = r_0 \equiv \lambda^2 \nu_0/C_0,
\eeq
where $C_0$ is the bare orthorhombic elastic constant.
At this point the \emph{renormalized} orthorhombic elastic constant
$\bar{C}_0 \equiv C_0 - \lambda^2 \nu_0/r$ vanishes, triggering a simultaneous orthorhombic instability.
We take $r_0$ to be a small parameter, i.e., the effective energy
scale generated by the coupling $\lambda$ is small compared to Fermi energy. Technically, this allows
to track how the properties of the familiar electron-only theory are
recovered at a sufficiently high temperature.

\textbf{Direction selective criticality.}
This is an inherent property of acoustic instabilities of a solid whereby criticality, or the vanishing of the
acoustic phonon velocity, is restricted to certain high-symmetry directions in the Brillouin zone such as
$\hat{q}_{1,2} \equiv (\hat{q}_x \pm \hat{q}_y)/\sqrt{2}$ for a tetragonal-orthorhombic transition~\cite{Cowley1976}.
Along the remaining directions the non-critical strains come into play.
This physics is well-known from studies of structural transitions~\cite{Larkin1969,Levanyuk1970,Villain1970},
and its relevance for the finite-$T$ structural/nematic transition in FeSC has also been pointed
out~\cite{Cano2010,Karahasanovic2016}. Our goal here is to study how this physics affects the metal's quantum critical
properties.

%%%%%%%%%%%%%%%%%%%%%%%%%%%%%%%%%%%%%%%%%%%%%%%%%%%%%%%%%%%%%%%%%%%%%%%%%%%%%%%%%%%%%%%%%%%%%%%%%%%%%%%%%%
\begin{figure}[!!t]
\begin{center}
\includegraphics[width=0.7\linewidth,trim=0 0 0 0]{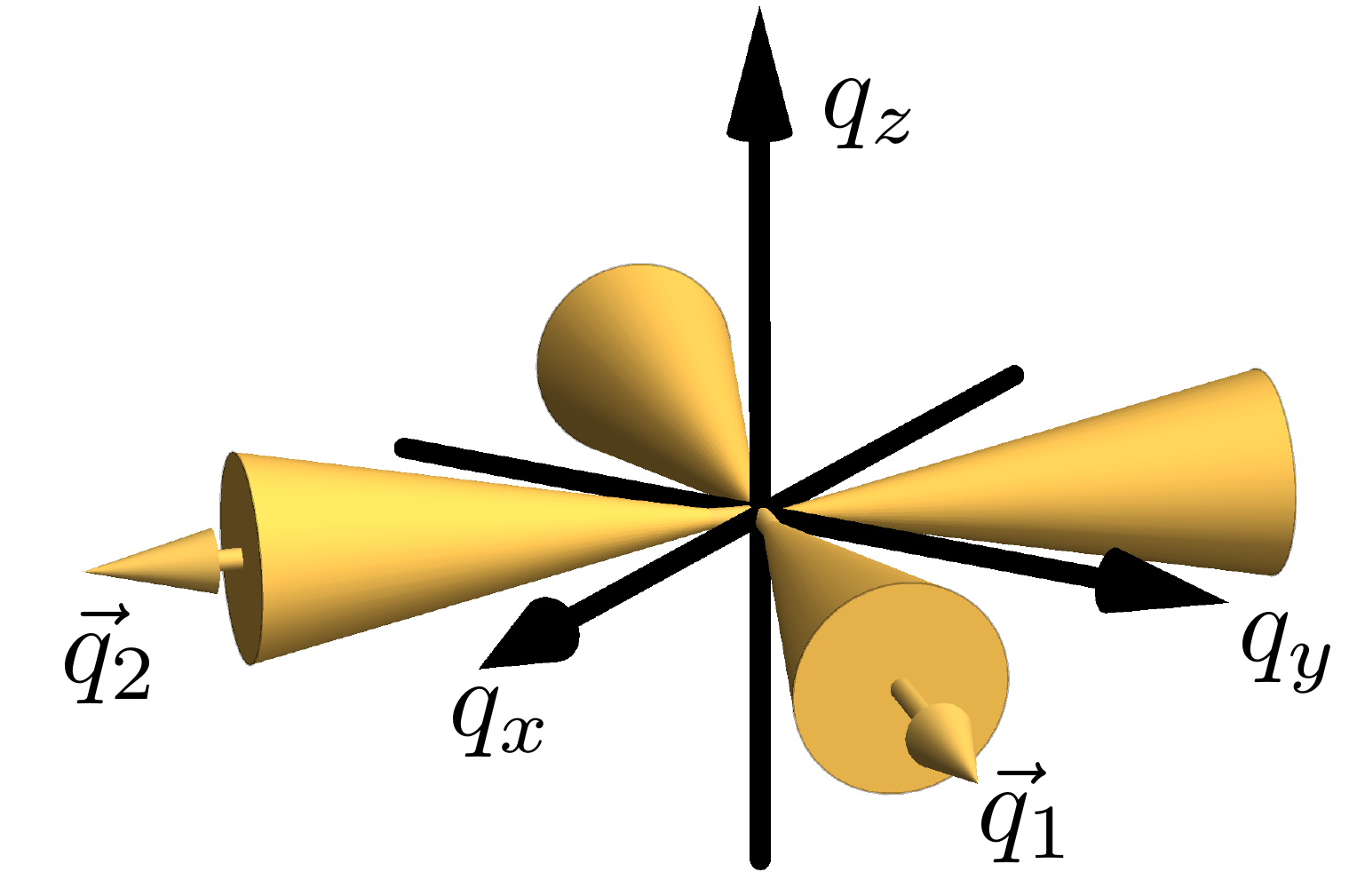}
\caption{
\textbf{Direction selective criticality.}
At the tetragonal-orthorhombic transition the critical directions in momentum space are restricted to
$\hat{q} \approx \pm \hat q_{1,2} = \pm (\hat q_x \pm \hat q_y)/\sqrt{2}$ as indicated by the yellow cones.
As a consequence, momentum scaling is anisotropic, see text.
}
\label{fig2}
\end{center}
\end{figure}
%%%%%%%%%%%%%%%%%%%%%%%%%%%%%%%%%%%%%%%%%%%%%%%%%%%%%%%%%%%%%%%%%%%%%%%%%%%%%%%%%%%%%%%%%%%%%%%%%%%%%%%%%%%%%

In the presence of the nemato-elastic coupling $\lambda$ the strain and the electron-nematic degree of freedom hybridize,
and the resulting mode inherits the above anisotropy.
The hybridization can be incorporated by integrating out the strain fluctuations
giving rise to a renormalization of the nematic susceptibility of Eq.~\eqref{eq:chi-n},
$\chi^{-1} = \chi^{-1}_0 - \Pi$, with
\beq
\label{eq:Pi}
\Pi(\bq, i\Omega_n) = \frac{\lambda^2}{\rho} \sum_{\mu}
\left( {\bf a}_{\bq} \cdot \hat u_{\bq, \mu} \right)^2/
\left( \omega^2_{\bq, \mu} + \Omega_n^2 \right).
\eeq
Here $\rho$ is the density, $\mu$ is the
polarization index, $ {\bf a}_{\bq} \equiv (q_x, -q_y,0)$,
and $\hat u_{\bq, \mu}$
is the polarization vector for the bare acoustic phonons
with angle-dependent velocity ${\bf v}^{(0)}_{\hat{q},\mu}$ and dispersion
$\omega_{\bq, \mu} = {\bf v}^{(0)}_{\hat{q},\mu} \cdot \bq$.
To lowest order in $r_0$, the frequency dependence of $\Pi$ can be dropped. Then
both the numerator and the denominator of $\Pi$ are $\mathcal{O}(q^2)$.
This implies that the effect of the nemato-elastic coupling is to soften the mass of the nematic fluctuations,
albeit with an angular dependence, i.e. $ r \rightarrow r(\hat{q}) \equiv r - \nu_0 \Pi(\bq \rightarrow 0, \Omega_n=0)$.
Note, $r(\hat{q})$  possesses the four-fold symmetry of the crystal lattice in the non-nematic phase.
As shown in the SI~\cite{supplemental},
an immediate consequence of this angular dependent mass is that criticality is restricted to the
two high-symmetry directions $\hat{q} = \pm \hat{q}_{1,2}$ only,
for which $r(\hat{q})=0$ at the QCP, see Fig.~\ref{fig2}. The remaining directions stay non-critical since
$r(\hat{q} \neq \hat{q}_{1,2}) > 0$ even at the QCP.

In the following we assume that all
the bare elastic constants are of order $C_0$, such that the entire lattice effect can be modeled by the single parameter
$r_0$.  With this simplification, that does not change the results qualitatively, the critical static nematic susceptibility
is given by
$
\chi^{-1} (\bq \approx \bq_1)
\propto r_0 (q_2^2 + q_z^2)/q_1^2 + q_1^2.
$
Note, the criticality around $\bq_2$ can be deduced by $q_1 \leftrightarrow q_2$.
This leads to two important conclusions. First, even if the electronic sub-system has two-dimensional dispersion,
as in the cuprates and the FeSC, the $q_z$ dependence of $\chi(\bq)$ is generated by
the lattice. Second, the direction selective
criticality leads to
anisotropic scaling with $(q_2, q_z) \sim q_1^2$ around $\bq_1$, see Fig.~\ref{fig2}.
Since each non-critical direction scales as twice the critical one,
this is equivalent to a theory with isotropic scaling in an enhanced effective space dimension
$d_{\rm eff} =5$~\cite{Cowley1976,Folk1976,Zacharias2015}.
Thus, the effect of fluctuations are weaker
compared to the electron-only theory.

%====================
\begin{figure}[!!t]
\begin{center}
\includegraphics[width=\linewidth,trim=0 0 0 0]{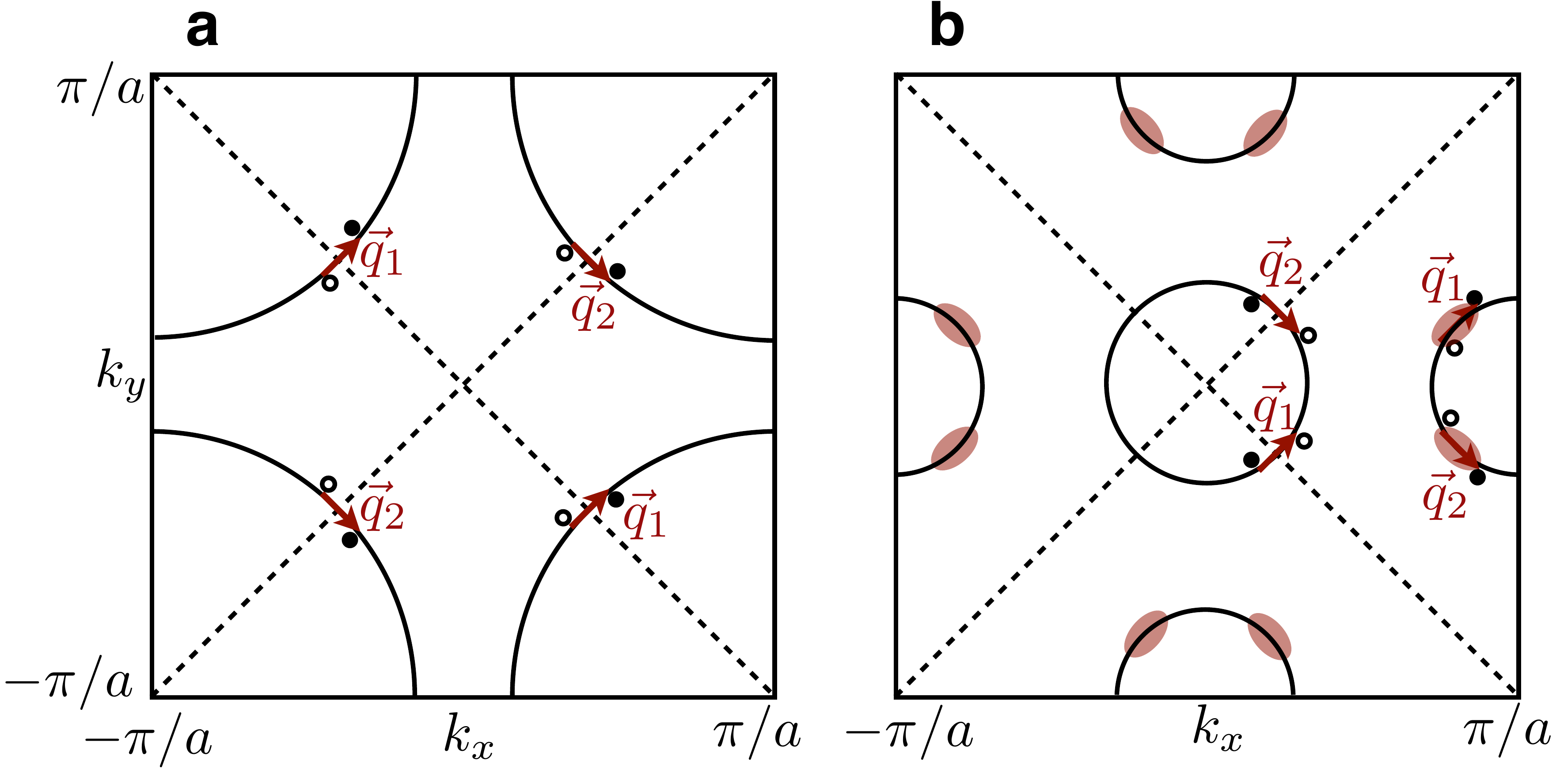}
\caption{
\textbf{Fermi surface dependent critical dynamics and the appearance of ``hot'' spots.}
Schematic Fermi surfaces of (a) the cuprates and (b) the iron based superconductors (FeSC).
The form factor accompanying the interaction between the electrons and the nematic boson
$h_{\bf{k}} \sim \cos k_x - \cos k_y =0$
along the dashed lines.
The critical bosons are restricted to the directions $\hat{q}_{1,2}$, see Fig.~(\ref{fig2}).
Landau damping is only possible via creation of particle-hole pairs at special points on the Fermi surfaces where
$\hat q_{1,2}$ is tangential, provided the form factor remains finite. This is the case only for the electron pockets
centered around $(\pi,0)$ and $(0, \pi)$ in (b).
Consequently, critical dynamics is ballistic in (a) and damped in (b) at the lowest energy. For the same reason
``hot'' spots with reduced fermion lifetimes (red patches) appear only on the electron pockets of (b). The
remaining Fermi surfaces stay ``cold''.
}
\label{fig3}
\end{center}
\end{figure}
%====================

\textbf{Fermi surface dependent dynamics.}
The effect of the lattice is indirect.
Since $\Pi$ is essentially static at small $r_0$, the critical dynamics
is generated by the excitation of particle-hole pairs in the Fermi sea,
%electrons,
and is given by $D(\bq, i \Om_n)$ of Eq.~(\ref{eq:chi-n}). In
electron-only theories this invariably leads to Landau-damping along generic directions $\hat{q}$,
and a dynamical exponent $z=3$.
However, with finite $\lambda$ the lattice imposes that $z$ is determined by $D(\bq \approx \bq_{1,2}, i \Om_n)$,
and the question is whether there is Landau-damping
along these directions. As we argue below, this depends on the Fermi surface,
leading to two different universality classes.

The important point is that the interaction between the nematic collective mode and the electrons, given by
$
\mathcal{H}_{\rm nem-el} \propto \sum_{\bq, \bk} h_{\bk}  c^{\dagger}_{\bk + \bq/2} c^\pdag_{\bk- \bq/2} \phi_{\bq}
$
in usual notations,
is invariably accompanied by a form factor $h_{\bk}$ that transforms as $(k_x^2-k_y^2)$.
Note, Landau damping requires electrons to scatter along the Fermi surface. This implies that the damping of
a collective mode with momentum along $\hat{q}$ depends on the form factor at those particular points
on the Fermi surface where $\hat{q}$ is tangential to the surface.

\emph{Ballistic nematicity.} Consider the Fermi surface of the cuprates, shown in Figure~\ref{fig3}(a).
The possibility of
Landau damping with bosonic momentum $\bq_1$ involve points on the Fermi surface which intersect with the
$k_x = - k_y$ dashed line, and along this
line the form factor $h_{\bk} =0$. Thus, there is no Landau damping, and we get
$D(\bq_1, i \Om_n) \propto \Om_n^2/(v_F q_1)^2$, leading to
ballistic critical dynamics at the lowest temperatures and frequencies, with dynamical
exponent $z=2$~\cite{Zacharias2009}.

\emph{Damped nematicity.} Now consider the typical Fermi surface of the FeSC with hole and electron pockets around the zone center,
and around $(\pi,0)$ and $(0,\pi)$, respectively, as shown in Figure~\ref{fig3}(b). For the same reason as above,
the hole pocket does not give rise to Landau damping of the critical mode.
But, since the centers of the electron pockets are shifted, $h_{\bk}$ is finite everywhere on
the electron Fermi surface, and the critical mode  gets damped.
This leads to the standard
$D(\bq_1, i \Om_n) \propto |\Om_n|/(v_F q_1)$ and exponent $z=3$. The damping only involves
certain hot spots of the electron pockets, on which we comment further below.
%====================
\begin{figure}[!!!!!!!t]
\begin{center}
\includegraphics[width=\linewidth,trim=0 0 0 0]{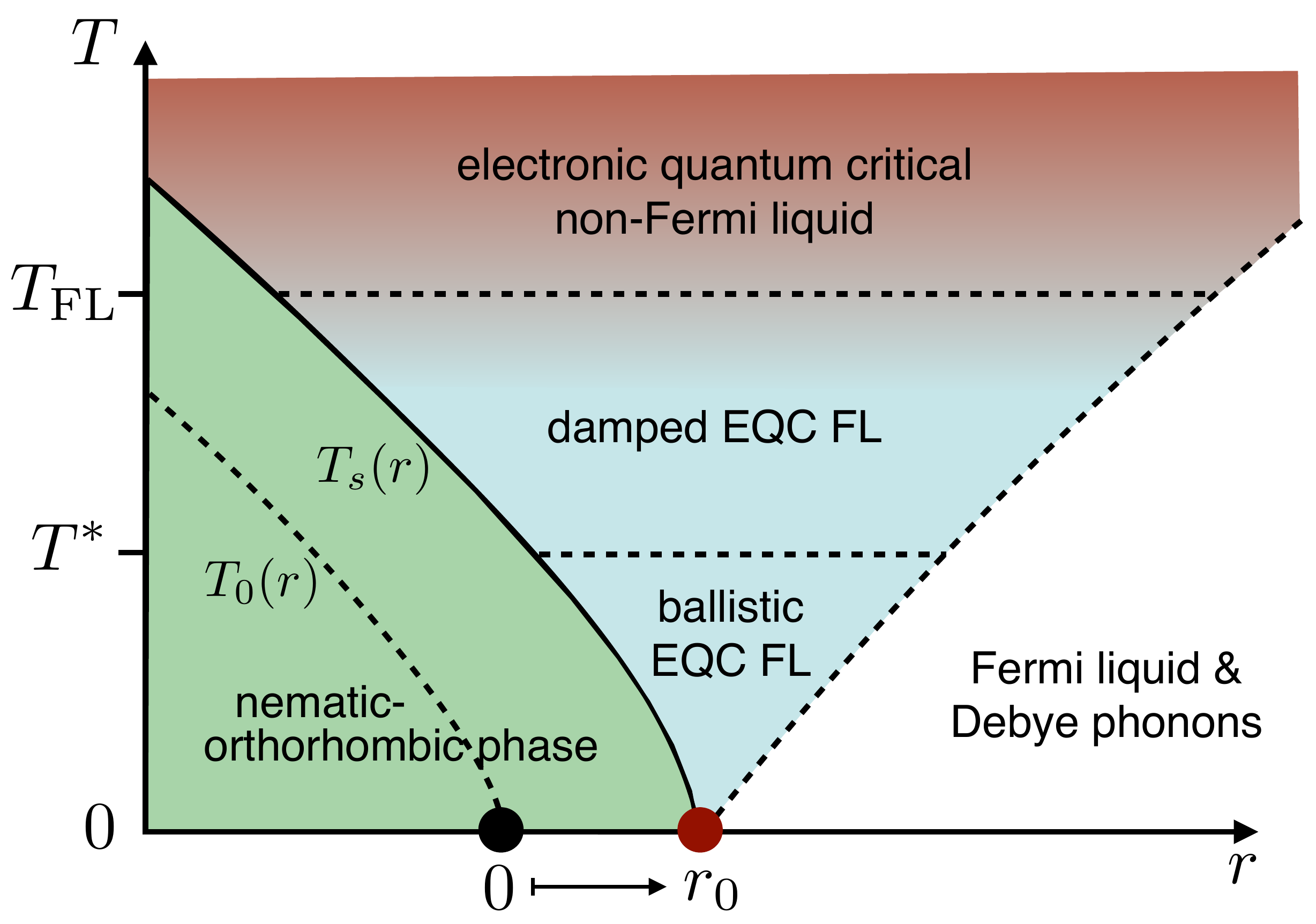}
\caption{
\textbf{Phase diagram with Ising-nematic quantum critical point (QCP).} $r$ is the control parameter.
Nemato-elastic coupling shifts the QCP from $r=0$ (black circle) to $r=r_0$ (red circle),
and the transition temperature from $T_0(r)$ to $T_s(r)$.
$r_0 \ll 1$
is the ratio between the lattice-generated energy scale and Fermi energy $E_F$. Above the
temperature scale $T_{\rm FL} \sim r_0^{3/2} E_F$ the nemato-elastic coupling can be neglected, and the familiar
electron-only theory of nematicity giving non Fermi liquid (NFL) physics is appropriate. $T_{\rm FL}$ is a crossover
to Fermi liquid physics. Below $T_{\rm FL}$ nemato-elastic coupling is important, and criticality is direction
selective (see Fig.~\ref{fig2}), as in elastic quantum criticality (EQC).
For the ballistic universality class, exemplified by the cuprates, there is an additional crossover at
$T^* \sim r_0^2 E_F$.  For the damped universality class,
exemplified by the iron superconductors, $T^* =0$. Their respective thermodynamics
are given by equations~(\ref{eq:gamma-ballistic}) and~(\ref{eq:gamma-damped}).
}
\label{fig4}
\end{center}
\end{figure}
%====================

\textbf{Critical thermodynamics.}
For the sake of concreteness henceforth we assume that the electronic dispersion is two-dimensional.
The free energy of the nematic fluctuations is $F = (T/2) \sum_{\bq, \Om_n} \log \chi^{-1} (\bq, i\Om_n)$, and
the critical phase diagram is summarized in Fig.~\ref{fig4}.
There are two important regions in $\bq$-space: $(i)$  $\hat{q} \approx \hat{q}_{1,2}$ (shaded area in Fig.~\ref{fig2}),
and $(ii)$
 $q_z \gg (q_1, q_2)$. For $(ii)$, the entire nemato-elastic coupling can be neglected, and we get the
 susceptibility of the electron-only theory with
$
\chi^{-1} \propto r_0 + q_{2d}^2 + |\Om_n|/q_{2d}$,
where $\bq_{2d} = (q_1, q_2)$.
Since it covers a larger volume in $\bq$-space, the contribution from $(ii)$ gives the leading term.
Thus, above the temperature scale $T_{\rm FL} \sim r_0^{3/2} E_F$, where $E_F$ is Fermi energy in temperature unit,
we recover the
usual electron-only theory with isotropic two-dimensional criticality and $\gamma(T) \propto 1/T^{1/3}$.
However, for $T \ll T_{\rm FL}$ this mode becomes massive giving Fermi-liquid type (FL) contribution
$\gamma(T) \propto 1/r_0^{1/2}$.
In this low $T$-regime the nemato-elastic coupling sets in, and direction selective criticality is
restricted to region $(i)$. The associated thermodynamics is as follows.

\emph{Ballistic nematicity (Cuprates).}
In this case
$
D (\bq \approx \bq_1, i\Om_n) \propto \Om_n^2/q_1^2 + (q_2/q_1)^2|\Om_n|/q_1,
$
where the last term indicates that Landau damping requires a finite $q_2$-component. The competition
between these two terms yields an additional crossover scale $T^*  \sim r_0^2 E_F$.
%give the crossover scale $T_1 \sim r_0^2 E_F$.
For $T \ll T^* $ the dynamics is ballistic,
giving the scaling $|\Om_n| \sim q_1^2$. But, above $T^* $ the dynamics is damped, with the scaling $|\Om_n| \sim r_0 q_1$.
In both these two regimes momentum scaling $(q_2, q_z) \sim q_1^2/r_0^{1/2}$ is anisotropic,
and the anisotropy extends up to the temperature $T_{\rm FL}$. The critical free energy can be estimated from the above scaling
(for detailed calculation see SI~\cite{supplemental}).
For $T \ll T_{\rm FL}$, including the non-critical contribution from $(ii)$ we get
\beq
\label{eq:gamma-ballistic}
\gamma (T) = \left\{\begin{array}{ll}
\displaystyle
\frac{c_1}{r_0^{1/2} E_F} + \frac{c_2 T^{3/2}}{r_0 E_F^{5/2}}, &  T \ll T^*, \\[1.5em]
\displaystyle
\frac{c_1}{r_0^{1/2} E_F} + \frac{c_3 T^4}{r_0^6 E_F^5}, &  T^* \ll T \ll T_{\rm FL}.
\end{array}
\right.
\eeq
Note, the $T^{3/2}$ contribution appears also in the context of elastic quantum criticality (EQC), even in the absence of
itinerant electrons~\cite{Zacharias2015}.

\emph{Damped nematicity (FeSC).}
In this case there is finite Landau damping even for $\hat{q} = \hat{q}_1$ so that
$D(\bq \approx \bq_1, i \Om_n) \propto |\Om_n|/(v_F q)$.
There is no physics related to the crossover $T^*$. For $T \ll T_{\rm FL}$ this leads to
\beq
\label{eq:gamma-damped}
\gamma (T) = \frac{c_1}{r_0^{1/2} E_F} + \frac{c_4 T^{2/3}}{r_0 E_F^{5/3}}, \quad  T \ll T_{\rm FL}.
\eeq

In the above $c_{1,\cdots,4}$ are numerical prefactors.
For both cases, once the nemato-elastic coupling sets in below $T_{\rm FL}$,
the leading thermodynamics is
Fermi liquid type, while the critical contribution is subleading, in stark contrast to what the electron-only theory predicts.
Note, in our theory we do not expect increased entropy in the nematic phase~\cite{gegenwart2006,rost2011}.

\textbf{Electron Self-energy.} We calculate at
zero temperature the frequency dependence of the
electron self-energy on the Fermi surface, i.e., $\Sigma_{\bk_F}(i \om_n) \propto h_{\bk_F}^2 S_{\hat{k}_F}(i \om_n)$,
where
$S_{\hat{k}_F}(i \om_n) = \int_{\bq, \nu_n} \chi(\bq, i \nu_n) G_{\bk_F + \bq}(i \om_n + i\nu_n)$, and $G$ is the
electron Green's function.
As in the free energy calculation, the regions $(i)$ and $(ii)$ of the $\bq$-space are important.
At sufficiently high frequency $|\om_n| \gg T_{\rm FL}$ the contribution from $(ii)$ gives $S(i\om_n) \propto |\om_n|^{2/3}$,
and the entire
Fermi surface is hot (barring the points where $h_{\bk_F}=0$). Thus, at high frequency we recover the properties of the
electron-only critical theory.
For low frequency this contribution turns into a non-critical Fermi liquid correction with $S(i\om_n) \sim -i \om_n/r_0^{1/2}$,
which guarantees that the real part of the self energy stays Fermi liquid type everywhere on the Fermi surface.

The contribution from region $(i)$ can lead to singular self-energy provided it involves electrons scattering parallel to the
Fermi surface. This implies that at most we expect ``hot spots'' where electronic lifetimes
are short, see Fig.~(\ref{fig3}). However, for the Fermi surface of the cuprates, as well as for the hole Fermi pockets
of the FeSC, the vanishing
form factor $h_{\bk}$ at these points imply that the nematic fluctuation induced hot spots do not survive (the familiar
hot-spot or Fermi-arc physics of the cuprates is presumably related to either spin fluctuations or Mott physics,
which are not treated here).
On the other hand, the
nematic fluctuation induced hot spots do survive
on the electron pockets of the FeSC for which $h_{\bk} \approx 1$. As shown in the SI~\cite{supplemental}, the region
$(i)$ gives a subleading critical contribution to self-energy $\Sigma(i \om_n)_{\rm cr} \propto |\om_n|^{4/3}/r_0^{1/2}$,
which leads to a reduced lifetime for electrons at these hot spots. The arc lengths of the spots scale as
$(|\om_n|^{1/3}/r_0^{1/2})k_F$, and thus their contribution to $\gamma (T) \propto T^{2/3}$, which is consistent
with Eq.~\eqref{eq:gamma-damped}.

\textbf{Discussion.} The nemato-elastic coupling in Eq.~(\ref{eq:nem-latt-coupling}) shifts not only the QCP, but also the
finite-$T$ transition from $T_0$ to $T_S$, see Fig.~(\ref{fig4}). Here $T_0$ is the nominal nematic transition temperature of
the electron subsystem in the absence of the coupling $\lambda$. Thus, the dimensionless parameter $r_0$ can
be estimated as $r_0 \sim (T_s - T_0)/E_F$.
Experimentally, $T_0$ is accessible from, say, electronic Raman scattering~\cite{Gallais2016}.
At present there is no clear experimental evidence of a nematic QCP in the cuprate phase diagram.
Consequently, iron-based superconductors are better suited to study effects of nemato-elastic coupling.
In BaFe$_2$As$_2$ we get $T_0 \sim$ 90 K~\cite{gallais2013}, and
$T_s \sim$ 138 K.
We estimate the Fermi temperature from the bottom of the electron bands as measured by photoemission, which are
around 50 meV in BaFe$_2$As$_2$~\cite{brouet2010}.
Thus, overestimating Fermi temperature $T_F \sim$ 1000 K, gives a conservative estimate of $r_0 \sim$ 0.05,
and  $T_{FL} \sim$ 10 K, or more, near the nematic QCP of Ba$($Fe$_{1-x}$Co$_x)_2$As$_2$.

For the iron-based superconductors we predict Fermi liquid behavior below $T_{FL}$
in thermodynamics and in single-particle properties, except at
``hot spots'' on the electron pockets where non canonical Fermi liquid behavior is expected.
Our predictions can be tested by photoemission, and by quasiparticle interference effects in tunneling spectroscopy
upon suppression of the superconducting phase in these systems.

We thank C. Max and A. Rosch for helpful discussions.
I.P. acknowledges financial support from ANR grant ``IRONIC'' (ANR-15-CE30-0025-01).
M.G. acknowledges support from SFB 1143 ``Correlated Magnetism: From Frustration To Topology''.

%%%%%%%%%% Merge with supplemental materials %%%%%%%%%%
\pagebreak
\widetext
\begin{center}
\textbf{\large Supplementary Material for ``Lattice effects on nematic quantum criticality in metals''}
\end{center}
%%%%%%%%%% Merge with supplemental materials %%%%%%%%%%
%%%%%%%%%% Prefix a "S" to all equations, figures, tables and reset the counter %%%%%%%%%%
\setcounter{equation}{0}
\setcounter{figure}{0}
\setcounter{table}{0}
\setcounter{page}{1}
\makeatletter
\renewcommand{\theequation}{S\arabic{equation}}
\renewcommand{\thefigure}{S\arabic{figure}}
%\renewcommand{\bibnumfmt}[1]{[S#1]}
%\renewcommand{\citenumfont}[1]{S#1}
%%%%%%%%%% Prefix a "S" to all equations, figures, tables and reset the counter %%%%%%%%%%
%%%%%%%%%%%%%%%%%%%%%%%%%%%%%%%%%%%%%%%%%%%%%%%%%%%%%%%%%%%%%%%%%%%%%%%%%%

\section{Nemato-elastic coupling and direction selective criticality}

In this section we provide the mathematical details of how nemato-elastic coupling leads to direction selective
criticality. We start with an explicit expression for the elastic energy, and then we discuss how the electronic
nematic susceptibility $\chi_0(\bq, i \Omega_n)$, see equation (1) of the main text, is renormalized
to $\chi(\bq, i \Omega_n)$ by the equation
\begin{align}
\label{eq:chi-norm}
\chi^{-1}(\bq, i \Omega_n) = \chi^{-1}_0(\bq, i \Omega_n) - \Pi(\bq, i \Omega_n)
\end{align}
in the presence of the
nemato-elastic coupling given by equation (2) of the main text. We derive the expression for $\Pi(\bq, i \Omega_n)$,
and we show how this leads to the concept of a four-fold symmetric mass, and, thus, to direction selective criticality.

\subsection{Elastic free energy \& normal modes}
The most general elastic free energy for a tetragonal system, to lowest order in the strains, is given
by~\cite{Landau-Lifshitz,Cowley1976}
\begin{align}
\label{eq:FE}
F_E &= \int d^3 \br \left[ \frac{c_{11}}{2} \left( \varepsilon_{xx}(\br)^2 + \varepsilon_{yy}(\br)^2 \right)
+ \frac{c_{33}}{2} \varepsilon_{zz}(\br)^2
+ 2 c_{44} \left( \varepsilon_{xz}(\br)^2 + \varepsilon_{yz}(\br)^2 \right) + 2 c_{66} \varepsilon_{xy}(\br)^2
\right. \nonumber \\
 &+ \left. c_{12} \varepsilon_{xx}(\br) \varepsilon_{yy}(\br)
+ c_{13} \left( \varepsilon_{xx}(\br) + \varepsilon_{yy}(\br) \right) \varepsilon_{zz}(\br) \right].
\end{align}
Here $\varepsilon_{ij}(\br) \equiv \varepsilon_{ij} + \frac{i}{2} \sum_{\bq \neq 0}
\left[ q_i u_j(\bq) + q_j u_i(\bq) \right] e^{i \bq \cdot \br}$, with $(i,j) = (x, y, z)$ are the local strains,
which has an uniform component $\varepsilon_{ij}$ and a fluctuating part that describe the acoustic phonons. The latter is
defined in terms of the atomic displacement ${\bf u}$, and $c_{11}$ etc.\ denote elastic
constants in Voigt notation. In particular, the local orthorhombic strain, that enters equation (2) of the main text,
is defined by $\varepsilon(\br) \equiv  \varepsilon_{xx}(\br) - \varepsilon_{yy}(\br)$, and the associated
elastic constant is $C_0 \equiv (C_{11} - C_{12})/2$. Note, in the theory the bare elastic medium is stable, and the
bare elastic constants are finite and temperature independent.

The dynamical matrix ${\bf N}(\bq)$ is defined by the relation
$F_E \equiv \sum_{{\bf q} \neq 0} u^{\ast}_i(\bq) N_{ij}(\bq) u_j(\bq)/2$,
where summation over repeated indices is implied. We get
$N_{xx} = c_{11} q_x^2 + c_{66} q_y^2 + 4 c_{44} q_z^2$,
$N_{yy} = N_{11} (q_x \leftrightarrow q_y)$,
$N_{zz} = c_{33} q_z^2 + c_{44} (q_x^2 + q_y^2)$,
$N_{xy} = (c_{12} + c_{66}) q_x q_y$,
$N_{xz} = (c_{13} + c_{44}) q_x q_z$,
$N_{yz} = N_{xz} (q_x \leftrightarrow q_y)$.
We write
\begin{align}
\label{eq:normal}
{\bf u}(\bq) = \sum_{\mu} U_{\bq, \mu} \hat{u}_{\bq, \mu},
\end{align}
where $\hat{u}_{\bq, \mu}$ are the polarization vectors, $U_{\bq, \mu}$ are the associated displacements at $\bq$, and
$\mu = (\alpha, \beta, \gamma)$ is the polarization index.
The eigenvalue equation ${\bf N}(\bq) \hat{u}_{\bq, \mu} = \rho \omega_{\bq, \mu}^2
\hat{u}_{\bq, \mu}$, with the mass density $\rho$, defines the bare phonon dispersions $\omega_{\bq, \mu}$.

\subsection{Nemato-elastic coupling \& renormalization of nematic susceptibility}
The electronic nematic variable $\phi(\br)$ is characterized by the susceptibility
$
\chi_0^{-1} (\bq, i \Omega_n) = \nu_0^{-1} \left[ r + q^2/k_F^2 + D(\bq, i \Om_n) \right],
$
in the electron-only theory, and is given in equation (1) of the main text. Here $D(\bq, i \Om_n) \propto |\Om_n|/(v_F q)$
denotes Landau damped dynamics, and a dynamical exponent $z=3$, which is standard in an electron-only theory for a quantum
critical point (QCP) where the instability is at $\bq =0$, such as a nematic one. In particular, the
above susceptibility implies a mean field Landau free energy density $f_L = \frac{r \nu_0}{2} \phi_0^2$ for the transition, where
$\phi_0 \equiv \phi_{\bq=0}$ is the electron nematic order parameter with dimension of energy.
In other words, the QCP is at $r=0$. In the nematic phase
the electronic dispersions along $\hat{x}$ and $\hat{y}$ directions are inequivalent, and this is manifested by a
Fermi surface which is $C_2$ symmetric rather than $C_4$ symmetric, see Fig.~(1a) in the main text.
Our goal is to study how the critical theory is modified by the symmetry-allowed nemato-elastic coupling
$
\mathcal{H}_{\rm nem-latt} = \lambda \int d^d \br \phi (\br) \varepsilon(\br),
$
where $\lambda$ is the coupling constant with dimension of energy.

At the mean field level the effect of $\lambda$ is to couple $\phi_0$ with the uniform orthorhombic strain
$\varepsilon \equiv \varepsilon_{xx} - \varepsilon_{yy}$,
such that the Landau free energy density for the transition is modified to
\begin{align}
\label{eq:FL}
f_L = \frac{r \nu_0}{2} \phi_0^2 + \frac{C_0}{2} \varepsilon^2 + \lambda \nu_0 \phi_0 \varepsilon.
\end{align}
This implies that the QCP is shifted to a positive value of $r= r_0 \equiv \lambda^2 \nu_0/C_0$. This is the content of
equation (3) in the main text. $r_0$ is the ratio of a lattice generated energy scale to Fermi energy, and it can be
taken as a small parameter of the theory.
Simultaneously, the renormalized orthorhombic elastic constant
\begin{align}
\label{eq:CO}
\bar{C}_0 \equiv C_0 - \lambda^2 \nu_0/r,
\end{align}
softens to zero at the QCP, and in the nematic phase the lattice is orthorhombic with $\varepsilon \neq 0$, see Fig.~(1c)
in the main text.
Note, since the remaining strains, defined in equation (\ref{eq:FE}), do not couple to $\phi$ at the
mean field level, they remain non-critical variables in the theory. That is, their corresponding elastic constants are not
renormalized by $\lambda$ from their bare values.

At the level of fluctuations the nemato-elastic coupling can be written as
\begin{align}
\label{eq:H-ne}
\mathcal{H}_{\rm nem-latt} = i \lambda \sum_{\bq \neq 0}  \bf{a}_{\bq} \cdot \bf{u}_{\bq} \phi_{- \bq},
\end{align}
where $ {\bf a}_{\bq} \equiv (q_x, -q_y,0)$. Expressing the displacements in terms of the normal modes given by
equation~(\ref{eq:normal}), and noting that the phonon Green's function is
$\langle U_{\bq, \mu} U_{\bq^{\prime}, \mu^{\prime}} \rangle =
\delta_{\bq \bq^{\prime}} \delta_{\mu \mu^{\prime}} \rho^{-1}/(\omega^2_{\bq, \mu} + \Omega_n^2)$, it is simple to
infer that the result of integrating out the lattice variables is to obtain equation~(\ref{eq:chi-norm}) with
\begin{align}
\label{eq:S-Pi}
\Pi(\bq, i\Omega_n) = \frac{\lambda^2}{\rho} \sum_{\mu} \frac{\left( \bf{a}_{\bq} \cdot \hat{u}_{\bq, \mu} \right)^2}
{\omega^2_{\bq, \mu} + \Omega_n^2}.
\end{align}

\subsection{Direction selective criticality}
The acoustic phonon dispersion is linear in momentum with a direction dependent bare velocity ${\bf v}^{(0)}_{\hat{q},\mu}$, i.e.,
$\omega_{\bq, \mu} = \bf{v}^{(0)}_{\hat{q},\mu} \cdot \bq$.
This implies that $\Pi(\bq, \Omega_n=0)$ is only a function of the two angles
$\hat{q}$. Thus, the main effect of the nemato-elastic coupling is to renormalize the mass of the nematic fluctuations, which is
isotropic in the electron-only theory, and which becomes a four-fold symmetric function of $\hat{q}$. In other words,
\begin{align}
\label{eq:r-q}
r \rightarrow r(\hat{q}) \equiv r - \nu_0 \Pi(\bq, \Omega_n=0).
\end{align}
Note, collective modes with direction dependent masses are well known from studies of certain structural phase transitions,
such as an uniaxial ferroelectric transition that involves long range dipolar interaction~\cite{Larkin1969},
and acoustic instabilities where the long range force is mediated by the shear modes of the solid~\cite{Levanyuk1970}.

Already at this point it is clear that at the QCP, defined by $r=r_0$, the renormalized mass $r(\hat{q})$ cannot vanish along all the
directions. This leads to the concept of direction selective criticality, see Figure~(2) of the main text.
As we argue below, only the two high symmetry directions
$\hat{q}_{1,2} \equiv (\hat{q}_x \pm \hat{q}_y)/\sqrt{2}$ become critical, while the remaining $\hat{q}$ stay non-critical,
i.e.,  at the QCP
\begin{align}
\label{eq:r-non-cr}
r(\hat{q} \neq \pm \hat{q}_{1,2}) > 0.
\end{align}

The evaluation of the renormalized mass $r(\hat{q})$ and the identification of the critical directions can be performed
simply by diagonalizing the 3 $\times$ 3 dynamical matrix $N_{ij}$. However, this is cumbersome and less insightful. Instead, we
will restrict the evaluation of $r(\hat{q})$ to only along the high symmetry directions, and this is sufficient to identify the
critical directions.

Thus, along the direction $\hat{q}_1$ the only lattice eigenmode that contributes to mass renormalization
has eigenvalue $\omega_{\hat{q}_1}^2 = (C_0/\rho) q_1^2$
and eigenvector $\hat{u}_{\hat{q}_1} = (1, -1, 0)/\sqrt{2}$.
This leads to $\Pi(\bq, \Omega_n=0; \hat{q}=\hat{q}_1) = \lambda^2/C_0$.
Similarly, along the direction $\hat{q}_x$ the contributing lattice eigenmode has eigenvalue
$\omega_{\hat{q}_x}^2 = (C_{11}/\rho) q_x^2$
and eigenvector $\hat{u}_{\hat{q}_x} = (1, 0, 0)$, and this leads to
$\Pi(\bq, \Omega_n=0; \hat{q}=\hat{q}_x) = \lambda^2/C_{11}$. Using tetragonal symmetry we also infer that
$\Pi(\hat{q}_1) = \Pi(\hat{q}_2)$, and $\Pi(\hat{q}_x) = \Pi(\hat{q}_y)$. On the other hand, along $\hat{q}_z$ we get $\Pi =0$.
The stability of the bare lattice ensures that $C_{11} > C_0$ (one of the Born stability criteria)~\cite{Born1954}.
Thus, among the high symmetry directions, the mass is the softest
along $\hat{q}=\pm \hat{q}_{1,2}$. Furthermore, at the QCP $r=r_0$, and we get $r(\pm \hat{q}_{1,2}) = 0$, while the mass along the
remaining high symmetry directions stay positive.
The physics of this phenomena is the following. Among the various independent strains only the orthorhombic strain
$\varepsilon(\br)$ is critical,
since the associated renormalized elastic constant softens to zero at the QCP, c.f., equation~(\ref{eq:CO}).
All the remaining strains, with positive elastic constants, are non-critical. Now, for $\bq \neq 0$, the electronic nematic
variable $\phi_{\bq}$ couples to one or more non-critical strain along $\hat{q} \neq \pm \hat{q}_{1,2}$, and this leads to a finite
mass $r(\hat{q})$. This includes the generic non high symmetry directions as well.
While along the special directions $\hat{q} = \pm \hat{q}_{1,2}$, $\phi_{\bq}$ triggers only the critical strain,
and, therefore, at the QCP $r(\pm \hat{q}_{1,2})=0$. This completes the argument that only
$\pm \hat{q}_{1,2}$ are critical directions, while the rest are non-critical.

From the above discussion one can infer that small deviations from a critical direction, say $\hat{q} = \hat{q}_1$, can be
expressed as $r(\hat{q} \approx \hat{q}_1) = c_2 (q_2/q_1)^2 + c_z (q_z/q_1)^2$, where $c_{2,z}$ are pre-factors
that depend on the bare elastic constants. Now, in this work we are interested only in the leading temperature and frequency
dependencies and the associated exponents of various quantities, and not their numerical prefactors. On the other hand, in a
typical metal, the bare elastic constants are all of the order of 10 GPa, and their ratios are simply numbers of order one, that
we are not interested to track in this work. Consequently, the calculation simplifies immensely if we assume that the bare
elastic constants are all of the order of $C_0$, such that the entire lattice effect can be modeled by the single parameter $r_0$.
It is easy to show that $c_{2,z} \propto r_0$.
With this simplification the asymptotic form of the renormalized static nematic susceptibility can be written as
\begin{align}
\label{eq:static-chi-full}
\chi^{-1} (\bq \approx \bq_1) = \nu_0^{-1} \left[ r_0 ( q_2^2 + q_z^2)/q_1^2 + q_1^2/k_F^2 \right].
\end{align}
Thus, direction selective criticality leads to anisotropic scaling with $(q_2, q_z) \sim q_1^2$, see Figure~(2d)
in the main text.
This is standard for critical elasticity~\cite{Cowley1976,Folk1976,Zacharias2015}.

\section{Fermi surface dependent dynamics}
At small $r_0$ the dynamics generated by the phonons can be ignored, and that of the nematic boson $\phi$ is entirely
generated by its interaction with the electrons. This interaction has the structure,
\begin{align}
\label{eq:H-nem-el}
\mathcal{H}_{\rm nem-el} \propto \sum_{\bq, \bk} h_{\bk}  c^{\dagger}_{\bk + \bq/2} c_{\bk - \bq/2} \phi_{\bq},
\end{align}
where $h_{\bk} \sim \cos k_x - \cos k_y$ is a form factor that transforms as $(k_x^2-k_y^2)$.
This leads to a nematic polarization of the form
\begin{align}
\label{eq:D}
D(\bq, i \Om_n) \propto - T \sum_{\bk, \om_n} h_{\bk}^2 G_{\bk}(i\om_n) G_{\bk + \bq}(i\om_n + i\Om_n),
\end{align}
where $G_{\bk}(i\om_n) = (i \om_n - \ep_{\bk})^{-1}$ is the electron Green's function.
For the sake of concreteness, from now on we assume that the electronic sub-system is two-dimensional, as in the
case of the iron based superconductors and the cuprates.
We simplify the evaluation of the above by assuming a two-dimensional circular Fermi surface
where $\psi_{\bf k} = \angle ({\bf q}, {\bf k})$ and $\hat k = {\bf k}/k$. Note, the qualitative conclusions
will not change for Fermi surfaces with crystalline anisotropy.
The low energy contribution to the dynamics of the collective mode can be turned into a Fermi surface integral of
the type
\[
D(\bq, i \Om_n) \propto -i \Om_n \int_0^{2 \pi} \frac{d \psi_{\bk}}{2 \pi} \frac{h_{\bk}^2}
{i\Om_n - v_F \bq \cdot \hat{k}}.
\]
The evaluation of the above is standard, and we get~\cite{Zacharias2009}
\begin{align}
\label{eq:D-cuprate}
D(\bq, i \Om_n) \propto \frac{|\Om_n|}{v_F q} \sin^2 2 \theta_{\bq} + 2 \frac{\Om_n^2}{(v_F q)^2} \cos 4 \theta_{\bq},
\end{align}
where $\cos \theta_{\bq} = \hat{q} \cdot \hat{q}_1$.

Due to direction selective criticality, we need to evaluate $D(\bq, i \Om_n)$ along the critical directions
$\hat{q}_{1,2}$, and this depends crucially whether $h_{\bk}$ itself vanishes along these directions.
As shown in Figure~(3) of the main text, this is indeed
the case for a typical cuprate Fermi surface and for that of the hole pockets in the iron superconductors.
Thus, for the cuprate Fermi surface the Landau damping vanishes along the critical directions for which
$\theta_{\bq} = 0, \pi/2$, and we get
\begin{align}
\label{eq:D-cuprate2}
D(\bq \approx \bq_1, i \Om_n)_{\rm cuprate} \propto \frac{\Om_n^2}{(v_F q_1)^2},
\end{align}
which gives \emph{ballistic critical dynamics}, and dynamical exponent $z=2$.
On the other hand, for the iron superconductors (FeSC), the typical Fermi surface is comprised of
hole pockets centred at $(0,0)$, and electron pockets centred at $(\pi,0)$ and $(0,\pi)$. For the same reason
as above, the hole pockets cannot provide Landau damping of the critical nematic fluctuations. But, on the
electron pockets the form factor $h_{\bk} \approx 1$, and in this case we get
\begin{align}
\label{eq:D-FeSC}
D(\bq \approx \bq_1, i \Om_n)_{\rm FeSC} \propto \frac{|\Om_n|}{v_F q_1}.
\end{align}
In other words, we get back standard Landau damping and exponent $z=3$.
Thus, depending on the underlying Fermi surfaces, we get two different classes of nematic quantum criticality.

\section{Critical thermodynamics}
In this section we give details of the calculation of the free energy of the nematic fluctuations.
The results are summarized in the phase diagram of Figure~(4) in the main text. The free energy is given by
\begin{align}
\label{eq:free-energy}
F = (T/2) \sum_{\bq, \Om_n} \log \chi^{-1} (\bq, i\Om_n).
\end{align}
As mentioned in the main text, there are two regions in the momentum space that are important.
(i) The region $\hat{q} \approx \pm \hat{q}_{1,2}$, where critical fluctuations survive once the nemato-elastic
coupling is significant.
(ii) The region
$q_z \gg (q_1, q_2)$ for which the entire nemato-elastic coupling can be neglected, and where we
get the susceptibility of the electron-only theory with
\begin{align}
\label{eq:chi-2d}
\chi^{-1} \propto r_0 + q_{2d}^2/k_F^2 + |\Om_n|/(v_F q_{2d}),
\end{align}
and $\bq_{2d} = (q_1, q_2)$.
Note, since (ii) spans a larger volume in momentum space than (i), the leading contribution to thermodynamics
is from (ii) \emph{at all temperatures}.

The contribution from (ii) is straightforward to evaluate. Above the temperature scale $T_{\rm FL} \sim r_0^{3/2} E_F$ we recover the
usual electron-only theory with isotropic two-dimensional criticality and $\gamma(T) \propto 1/T^{1/3}$.
However, for $T \ll T_{\rm FL}$ it behaves as a massive mode, giving Fermi liquid type contribution with
$\gamma(T) \propto 1/r_0^{1/2}$.
In this low $T$-regime the nemato-elastic coupling sets in, and criticality is direction selective which is
restricted to region (i) of $\bq$-space. The associated thermodynamics now depends on the type of the dynamics of the
fluctuations, and, therefore, on the Fermi surface of the electrons.

\subsection{Ballistic nematicity (Cuprates)}
Combining equations (\ref{eq:static-chi-full}) and (\ref{eq:D-cuprate}), the critical nematic
susceptibility for $T \ll T_{\rm FL}$ is
\begin{align}
\label{eq:chi-cuprate}
\chi^{-1} (\bq \approx \bq_1, i\Om_n) = \nu_0^{-1}
\left[ r_0 ( q_2^2 + q_z^2)/q_1^2 + q_1^2/k_F^2 + \Om_n^2/(v_F q_1)^2 + (q_2/q_1)^2|\Om_n|/(v_F q_1) \right].
\end{align}
This leads to the following two regimes.

\emph{Low temperature regime $T \ll T^* \sim r_0^2 E_F$.} Here the dynamics is ballistic and the last term in the
above equation can be neglected. This leads to the scaling $|\Om_n| \sim q_1^2$, and $(q_2, q_z) \sim q_1^2/r_0^{1/2}$.
The critical free energy is given by
\[
F(T) = T \int \frac{dq_1 dq_2 dq_z}{(2\pi)^3} \ln \left( 1 - e^{-E_{\bq}/T} \right),
\]
where $E_{\bq}^2/E_F^2 = r_0(q_2^2 + q_z^2) + q_1^4$. Note, the above momentum integral is restricted to
$(q_2, q_z) \lesssim q_1^2/r_0^{1/2}$ where scaling holds.
To leading order we get
\begin{align}
\label{eq:F1-ballistic}
F(T) = - \frac{a_1}{(2\pi)^2}\frac{T^{3.5}}{r_0 E_F^{2.5}},
\end{align}
where $a_1 \approx - \int_0^{\infty} dq q^4 \int_0^1 dp p \ln \left( 1 - e^{-q^2 \sqrt{1 + p^2}} \right) = 0.24$.
Note, the numerical prefactor $a_1$ depends on the ultraviolet cutoff of the scaling  $(q_2, q_z) \sim q_1^2/r_0^{1/2}$,
i.e., on the upper cutoff of the $p$-integral. As such, it is a non-universal quantity whose order of magnitude is
significant, rather than its precise value.
It translates into a specific heat  coefficient $\gamma(T) \propto T^{3/2}$.
The $T^{3/2}$ contribution to $\gamma(T)$ also appears in the context of quantum critical elasticity even in the
absence of itinerant electrons~\cite{Zacharias2015}.
Note, this
critical contribution is only a subleading term, the leading one being the Fermi liquid type
$\gamma(T) \propto 1/r_0^{1/2}$ contribution of the region (ii).

\emph{Intermediate temperature regime $T^* \ll T \ll T_{\rm FL}$.} Here the last term in equation~(\ref{eq:chi-cuprate})
wins over the
ballistic $\Om_n^2/(v_F q_1)^2$ term, and the dynamics is damped. This leads to the scaling
$|\Om_n| \sim r_0 q_1$ and $(q_2, q_z) \sim q_1^2/r_0^{1/2}$, and a critical free energy
\begin{align}
\label{eq:F2-ballistic}
F(T) &= - \frac{1}{\pi^4} \int_0^{\om/(E_F r_0)} dq_1 \int_0^{q_1^2/r_0^{1/2}} dq_2 dq_z \int_0^{\infty} d \om
n_B(\om) \tan^{-1} \left( \frac{\om q_2^2/(q_1 E_F)}{q_1^4 + r_0q_2^2 + r_0 q_z^2} \right)
\nonumber \\
&= - 2 a_2 a_3 \frac{T^6}{r_0^6 E_F^5},
\end{align}
where $n_B$ is the Bose function,
$a_2 = \int_0^1 dq q^4 \int_0^1 dp p \int_0^{2\pi} d \phi/(2\pi) \tan^{-1} \left( \frac{p^2 \cos^2 \phi}{q(1+p^2)}\right)
\approx 0.02$, and
$ a_3 = \int_0^{\infty} (dx/\pi^3) x^5/(e^x-1) \approx 4$.
Note, $a_2$ depends on the ultraviolet cutoff of the momentum integrals, and is non-universal.
This gives a $\gamma(T) \propto T^4$, which is a rather weak
temperature dependence that is indistinguishable from higher order analytic Fermi liquid corrections in powers of $T^2$.
As before, the leading $T$ dependence in this regime is the Fermi liquid type
$\gamma(T) \propto 1/r_0^{1/2}$ contribution of the region (ii).

\subsection{Damped nematicity (FeSC)}
In this case the critical fluctuations stay damped down to the lowest temperatures, and $T^*$ can be set to zero.
Combining equations (\ref{eq:static-chi-full}) and (\ref{eq:D-FeSC}), the critical nematic
susceptibility for $T \ll T_{\rm FL}$ is
\begin{align}
\label{eq:chi-FeSC}
\chi^{-1} (\bq \approx \bq_1, i\Om_n) = \nu_0^{-1}
\left[ r_0 ( q_2^2 + q_z^2)/q_1^2 + q_1^2/k_F^2 + |\Om_n|/(v_F q_1) \right].
\end{align}
This gives the scaling
$|\Om_n| \sim q_1^3$ and $(q_2, q_z) \sim q_1^2/r_0^{1/2}$, and a critical free energy
\begin{align}
\label{eq:F-damped}
F(T) &= - \frac{1}{\pi^4} \int_0^{(\om/(E_F)^{1/3}} dq_1 \int_0^{q_1^2/r_0^{1/2}} dq_2 dq_z \int_0^{\infty} d \om
n_B(\om) \tan^{-1} \left( \frac{(\om q_1)/E_F)}{q_1^4 + r_0q_2^2 + r_0 q_z^2} \right)
\nonumber \\
&= -  a_4 a_5 \frac{T^{8/3}}{r_0 E_F^{5/3}},
\end{align}
$a_4 = \int_0^1 dq q^4 \int_0^1 dp p \tan^{-1} \left( \frac{1}{q^3(1+p^2)}\right)
\approx 0.09$, and
$ a_5 = (2/\pi^3) \int_0^{\infty} dx x^{5/3}/(e^x-1) \approx 0.12$. As in the earlier cases, the pre-factor $a_4$ is non-universal.
This leads to a critical $\gamma(T) \propto T^{2/3}$ which is subleading to the Fermi liquid contribution from region (ii).

This completes the demonstration that, for both the universality classes, below the scale $T_{\rm FL}$ the leading thermodynamics
is Fermi liquid type.

%====================
\begin{figure}[!!t]
\begin{center}
\includegraphics[width=0.5\linewidth,trim=0 0 0 0]{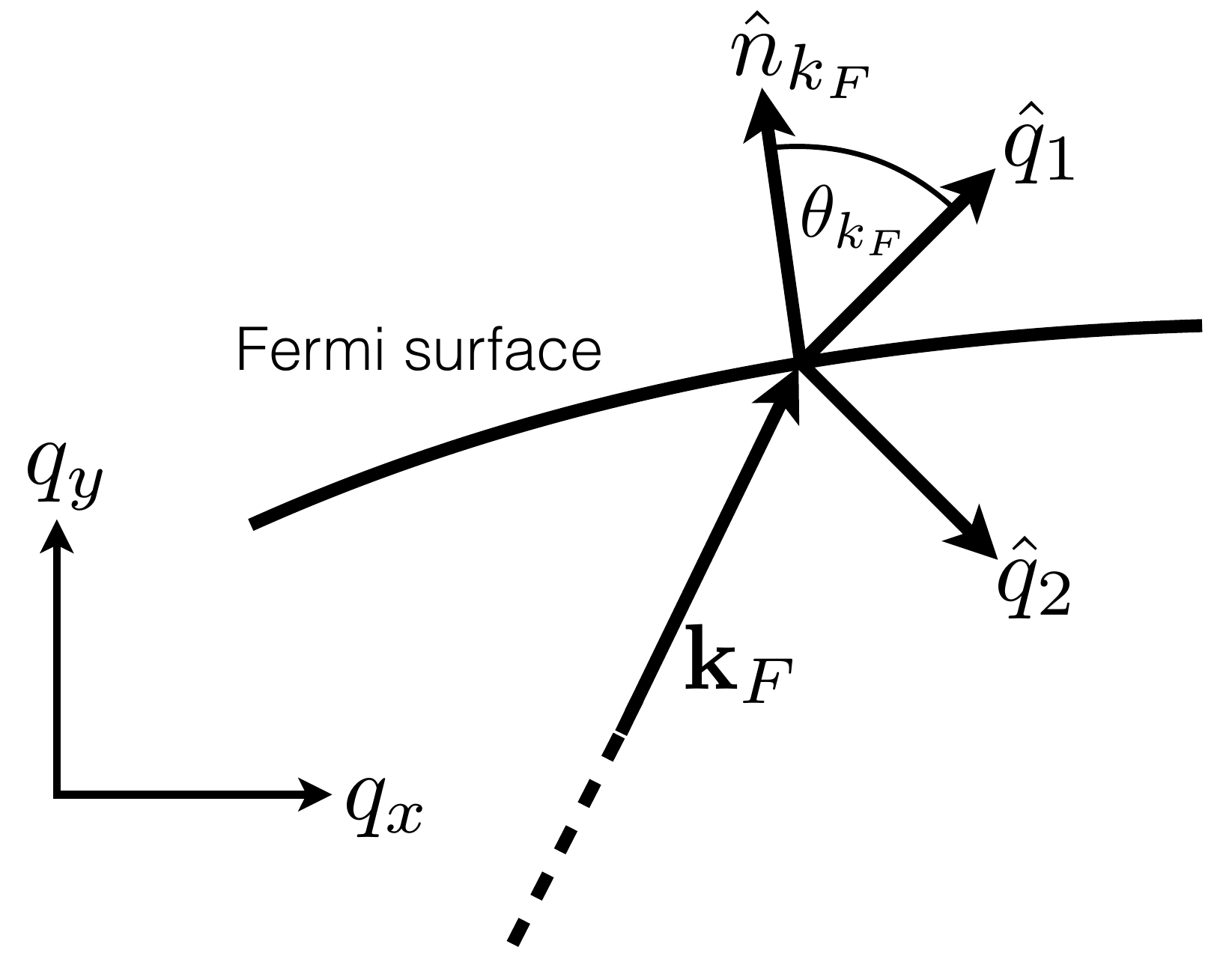}
\caption{
\textbf{A local patch of a Fermi surface.}
$\hat{n}_{k_F}$ is the direction normal to the Fermi surface at $\bk_F$. $\theta_{k_F}$ is the angle between $\hat{n}_{k_F}$
and the critical direction $\hat{q}_1 \equiv (\hat{q}_x + \hat{q}_y)/\sqrt{2}$.
}
\label{figS1}
\end{center}
\end{figure}
%====================

\section{Electron self-energy}
In this section we give the details of the calculation of the electron self-energy due to scattering with the critical
nematic fluctuations. This can be written as
\begin{align}
\label{eq:Sigma-general}
\Sigma_{\bk}(i \om_n) \propto h_{\bk}^2 S_{\bk}(i \om_n),
\end{align}
where
\begin{align}
\label{eq:S-fn}
S_{\bk}(i \om_n) = \int_{\bq, \nu_n} \chi(\bq, i \nu_n) G_{\bk + \bq}(i \om_n + i\nu_n).
\end{align}
In the following we calculate the frequency dependence of the self-energy at zero temperature
for electrons on the Fermi surface.
As in the case of the free energy calculation, there are two regions of $\bq$-space that are important,
namely (i) $\hat{q} \approx \hat{q}_{1,2}$, and (ii) $q_z \gg (q_1, q_2)$. The contribution from (ii) is straightforward.
At sufficiently high frequency $|\om_n| \gg T_{\rm FL}$ the mass term in equation~(\ref{eq:chi-2d}) can be neglected and
we recover the usual electron-only critical theory with $S^{(ii)}(i\om_n) \propto |\om_n|^{2/3}$, and the entire Fermi surface is
hot (except at points where $h_{\bk} =0$)~\cite{Lohneysen2007,Garst2010}.
For low frequency $|\om_n| \ll T_{\rm FL}$, the mass cannot be neglected, and we get a Fermi liquid type correction with
$S^{(ii)}(i\om_n) = -i \pi {\rm Sgn}(\om_n) \left( \pi |\om_n|/(2 r_0^{1/2})
- \om_n^2 \log (r_0^{3/2}/|\om_n|)/(2 r_0^2) \right)$.
The first term gives ${\rm Re} \Sigma \propto \om/r_0^{1/2}$, and the second term
${\rm Im} \Sigma \propto \om^2 \ln|\om|/r_0^2$, as is expected for a Fermi liquid in two space dimensions~\cite{Chaplik1971}.

In the low frequency range $|\om_n| \ll T_{\rm FL}$ we need to consider the contribution from region (i) of the $\bq$-space.
Note, the self-energy is expected to be singular only if the boson momentum is parallel to the Fermi surface. Since criticality
in this frequency range is restricted to only $\hat{q} \approx \pm \hat{q}_{1,2}$, this implies
that at most we expect ``hot spots'' around
$\hat{n}_{k_F} = \pm \hat{q}_{1,2}$, where $\hat{n}_{k_F}$ defines the direction normal to the Fermi surface
at $\bk_F$, see Figure~(\ref{figS1}).
The actual presence/absence of the hot spots depend on whether the form factor $h_{\bk}$ is finite
or if it vanishes at these points on the Fermi surface, see Figure~(3) of the main text. This contribution is discussed below.

\subsection{Ballistic nematicity (Cuprates)}
In this case the critical fluctuations have ballistic dynamics at low enough frequencies. The general expression for the
self-energy correction from region (i) is
\begin{align}
\label{eq:S-ballistic-1}
S^{(i)}_{\hat{n}_{k_F}}(i \om_n) = \int_{\bq, \nu_n} \frac{q_2^2}{q_2^4 + r_0(q_1^2 + q_z^2) + \nu_n^2} \frac{1}{i\om_n + i\nu_n
- q_1 \cos \theta_{k_F} - q_2 \sin \theta_{k_F}} + (\theta_{k_F} \rightarrow \theta_{k_F} + \pi/2),
\end{align}
where $\cos \theta_{k_F} \equiv \hat{n}_{k_F} \cdot \hat{q}_1$, see Figure~(\ref{figS1}).
The two terms above are contributions from $\hat{q} \approx \hat{q}_{2,1}$.
The above integral has contributions both from the fermion and the boson poles. For $\hat{n}_{k_F} \approx \pm \hat{q}_{1,2}$,
both the leading and the subleading contribution  is from the fermion pole. In the case of generic $\hat{n}_{k_F}$ only the
subleading term is from the boson pole, which we do not evaluate here. In order to estimate the fermion pole
contribution we introduce the orthogonal variables $\xi = q_1 \cos \theta_{k_F} + q_2 \sin \theta_{k_F}$ and
$\eta = - q_1 \sin \theta_{k_F} + q_2 \cos \theta_{k_F}$. We ignore the $\xi$ dependence of the boson propagator
and we get after the $\xi$ and $q_z$ integrals
\begin{align}
\label{eq:S-ballistic-2}
S^{(i)}_{\hat{n}_{k_F}}(i \om_n) \approx -i \frac{{\rm Sgn}(\om_n)}{r_0^{1/2}} \int_0^{|\om_n|} d \nu_n \int_0^1 d \eta \left[
\frac{\eta^2 \cos^2 \theta_{k_F}}{\sqrt{\eta^4 \cos^4 \theta_{k_F} + r_0 \eta^2 \sin^2 \theta_{k_F} + \nu_n^2}}
+ (\theta_{k_F} \rightarrow \theta_{k_F} + \pi/2) \right].
\end{align}
For the leading frequency dependence we can set $\nu_n =0$ in the integrand. This gives
\begin{align}
\label{eq:S-ballistic-3}
S^{(i)}_{\hat{n}_{k_F}}(i \om_n)_{\rm leading} = \frac{-i \om_n}{r_0^{1/2}} f(\theta_{k_F}),
\end{align}
where
\[
f(\theta_{k_F}) = \int_0^1 d \eta \frac{\eta \cos^2 \theta_{k_F}}{\sqrt{\eta^2 \cos^4 \theta_{k_F}
+ r_0 \sin^2 \theta_{k_F}}} + (\theta_{k_F} \rightarrow \theta_{k_F} + \pi/2).
\]
This implies that the contribution of region (i) to the real part of the self-energy is Fermi liquid like, and it is of the same
order as the contribution from region (ii). In order to obtain ${\rm Im} \Sigma$ we also study the subleading frequency
dependence of $S^{(i)}_{\hat{n}_{k_F}}(i \om_n)$.
For $\hat{n}_{k_F} = \pm \hat{q}_1$ the important contribution is from the first term
in equation~(\ref{eq:S-ballistic-2}) that involves the critical boson with momentum oriented
along $\hat{q}_2$, and vice versa.
We expand the integrand in $\nu_n$, and for the self-energy around
$\hat{n}_{k_F} = \pm \hat{q}_1$ we get
\[
S^{(i)}_{\hat{n}_{k_F}}(i \om_n; \theta_{k_F} \ll 1)_{\rm subleading} = \frac{i {\rm Sgn} (\om_n)}{r_0^{1/2}}
\int_0^{|\om_n|} d \nu_n \nu_n^2 \int_{\nu_n^{1/2}}^1 d \eta \frac{1}{\eta (\eta^2 + r_0 \theta_{k_F}^2)^{3/2}}.
\]
Note, the $\eta$-integral is now infrared divergent and it needs a suitable lower cutoff. This leads to
\beq
\label{eq:S-ballistic-4}
\begin{split}
S^{(i)}_{\hat{n}_{k_F}}(i \om_n; \theta_{k_F} \ll 1)_{\rm subleading}
&= \frac{i {\rm Sgn} (\om_n) |\om_n|^{3/2}}{r_0^{1/2}}, \quad \theta_{k_F} \ll \frac{|\om_n|^{1/2}}{r_0^{1/2}},
\\
&= \frac{-i {\rm Sgn} (\om_n) |\om_n|^{3}}{r_0^2 \theta_{k_F}^3} \log|\om_n|, \quad
\theta_{k_F} \gg \frac{|\om_n|^{1/2}}{r_0^{1/2}}.
\end{split}
\eeq
It is important to note that the form factor $h_{\bk} =0$ for $\theta_{k_F} = 0$. Thus,
in the above
the $|\om_n|^{3/2}$ self-energy term is accompanied by a vanishing pre-factor
$h_{\bk}^2 \propto \theta_{k_F}^2$. Using four-fold symmetry of the self-energy, the same argument is valid for
$\hat{n}_{k_F} = \pm \hat{q}_2$.
This guarantees that the leading ${\rm Im} \Sigma$ is from region (ii) and is Fermi liquid
like. \emph{Thus, with a cuprate type of Fermi surface the putative hot spots at
$\hat{n}_{k_F} = \pm \hat{q}_{1,2}$ are rendered cold by the vanishing form factor.}
Furthermore, one can show that the subleading term for generic $\hat{n}_{k_F}$ has the form
$S^{(i)}_{\hat{n}_{k_F}}(i \om_n; \theta_{k_F} \approx 1)_{\rm subleading} \propto |\omega_n|^{5/2}/r_0$
coming from the boson pole. This leads to a free energy $F(T) \propto T^{7/2}$, which is consistent with
equation~(\ref{eq:F1-ballistic}).

\subsection{Damped nematicity (FeSC)}
In this case the critical fluctuations have standard Landau damped dynamics since they couple to the particle-hole
continuum of the electron pockets. We get
\begin{align}
\label{eq:S-damped-1}
S^{(i)}_{\hat{n}_{k_F}}(i \om_n) = \int_{\bq, \nu_n} \frac{q_2^2}{q_2^4 + r_0(q_1^2 + q_z^2) + |\nu_n| q_2} \frac{1}{i\om_n + i\nu_n
- q_1 \cos \theta_{k_F} - q_2 \sin \theta_{k_F}} + (\theta_{k_F} \rightarrow \theta_{k_F} + \pi/2).
\end{align}
This can be evaluated as in the above. We introduce the variables $(\xi, \eta)$, and after the $\xi$ and $q_z$ integrals we get
\begin{align}
\label{eq:S-damped-2}
S^{(i)}_{\hat{n}_{k_F}}(i \om_n) \approx -i \frac{{\rm Sgn}(\om_n)}{r_0^{1/2}} \int_0^{|\om_n|} d \nu_n \int_0^1 d \eta \left[
\frac{\eta^2 \cos^2 \theta_{k_F}}{\sqrt{\eta^4 \cos^4 \theta_{k_F} + r_0 \eta^2 \sin^2 \theta_{k_F} + \nu_n \eta \cos \theta_{k_F}}}
+ (\theta_{k_F} \rightarrow \theta_{k_F} + \pi/2) \right].
\end{align}
The leading frequency dependence is Fermi liquid like as in the ballistic case, and is also given by equation~(\ref{eq:S-ballistic-3}).
The information about ${\rm Im} \Sigma$ is, however, in the subleading term, and we get
\[
S^{(i)}_{\hat{n}_{k_F}}(i \om_n; \theta_{k_F} \ll 1)_{\rm subleading} = \frac{i {\rm Sgn} (\om_n)}{r_0^{1/2}}
\int_0^{|\om_n|} d \nu_n \nu_n \int_{\nu_n^{1/3}}^1 d \eta \frac{1}{(\eta^2 + r_0 \theta_{k_F}^2)^{3/2}}.
\]
This gives
\beq
\label{eq:S-damped-3}
\begin{split}
S^{(i)}_{\hat{n}_{k_F}}(i \om_n; \theta_{k_F} \ll 1)_{\rm subleading}
&= \frac{i {\rm Sgn} (\om_n) |\om_n|^{4/3}}{r_0^{1/2}}, \quad \theta_{k_F} \ll \frac{|\om_n|^{1/3}}{r_0^{1/2}},
\\
&= \frac{i {\rm Sgn} (\om_n) |\om_n|^{2}}{r_0^{3/2} \theta_{k_F}^2}, \quad
\theta_{k_F} \gg \frac{|\om_n|^{1/3}}{r_0^{1/2}}.
\end{split}
\eeq
Note, on the electron pockets of the FeSC the form factor $h_{\bk} \approx 1$. Thus, the hot spots at
$\hat{n}_{k_F} = \pm \hat{q}_{1,2}$ survive on the electron pockets, where the single particle lifetime
is shorter than what is standard for Fermi liquids in two space dimensions
with ${\rm Im} \Sigma \propto \om^{4/3}$. Furthermore, the sizes of the hot spots
can be estimated as $\Delta \theta_{\rm hot} \sim |\om_n|^{1/3}/r_0^{1/2}$.
Note, the free energy associated with the hot spots can be estimated as $F(T) \propto T^{8/3}$, which is consistent with
equation~(\ref{eq:F-damped}). On the other hand, as in the
case of the cuprates, the hot spots do not survive on the hole pockets due to vanishing form factor.

Overall, we conclude that the effect of the nemato-elastic coupling is quite drastic on the single electron properties.
In an electron-only theory of nematic QCP we expect the Fermi surface to be isotropically hot with non Fermi liquid features.
In contrast, once the nemato-elastic coupling is included, Fermi liquid, or well-defined quasiparticles survive
at the lowest frequencies. Depending on the Fermi surface of the system at most there are hot spots where the quasiparticle
lifetime is shorter than what is usual in two dimensions. Thus, the entire cuprate Fermi surface stay cold,
while for the FeSC hot spots exist only on the electron pockets but not on the hole pockets.


\begin{thebibliography}{99}

\bibitem{Fradkin2010}
E. Fradkin, S. A. Kivelson, M. J. Lawler, J. P. Eisenstein, and A. P. Mackenzie, Annu. Rev. Condens. Matter
Phys. {\bf 1}, 153, (2010).

\bibitem{Nie2013}
L. Nie, G. Tarjus, and S. A. Kivelson, Proc. Natl. Acad. Sci. (USA) {\bf 111}, 7980 (2014).

\bibitem{Achkar2016}
A. J. Achkar, M. Zwiebler, C. McMahon, F. He, R. Sutarto, I. Djianto, Z. Hao,
M. J. P. Gingras, M. Hücker, G. D. Gu, A. Revcolevschi, H. Zhang, Y.-J. Kim, J. Geck, and D. G. Hawthorn,
Science {\bf 351}, 576 (2016).

\bibitem{Borzi2007}
R. A. Borzi, S. A. Grigera, J. Farrell, R. S. Perry, S. J. S. Lister, S. L. Lee, D. A. Tennant, Y. Maeno, and A. P. Mackenzie,
Science {\bf 315}, 214 (2007).

\bibitem{Lilly1999}
M. P. Lilly, K. B. Cooper, J. P. Eisenstein, L. N. Pfeiffer, and K. W. West,
Phys. Rev. Lett. {\bf 82}, 394 (1999).

\bibitem{Chu2010}
J.-H. Chu, J. G. Analytis, K. De Greve, P. L. McMahon, Z. Islam, Y. Yamamoto, and I. R. Fisher,
Science {\bf 329}, 824 (2010).

\bibitem{Fernandes2014}
R. M. Fernandes, A. V. Chubukov, and J. Schmalian, Nature Phys. {\bf 10}, 97 (2014).

\bibitem{Gallais2016}
Y. Gallais and I. Paul,
C. R. Phys. {\bf 17}, 113-139 (2016).

\bibitem{Johnston2010}
D. C. Johnston, Adv. Phys. {\bf 59}, 803 (2010).

\bibitem{Lohneysen2007}
H. v. L\"{o}hneysen, A. Rosch, M. Vojta, and P. W\"{o}lfle,
Rev. Mod. Phys. {\bf 79}, 1015 (2007).

\bibitem{Oganesyan2001}
V. Oganesyan, S. A. Kivelson, and E. Fradkin,
Phys. Rev. B {\bf 64}, 195109 (2001)

\bibitem{Metzner2003}
W. Metzner, D. Rohe, and S. Andergassen,
Phys. Rev. Lett. {\bf 91}, 066402 (2003)

\bibitem{Garst2010}
M. Garst and A. V. Chubukov,
Phys. Rev. B {\bf 81}, 235105 (2010).

\bibitem{Lee2009}
S.-S. Lee, Phys. Rev. B {\bf 80}, 165102 (2009).

\bibitem{Metlitski2010}
M. A. Metlitski and S. Sachdev, Phys. Rev. B {\bf 82}, 075127 (2010).

\bibitem{Mross2010}
D. F. Mross, J. McGreevy, H. Liu, and T. Senthil,
Phys. Rev. B {\bf 82}, 045121 (2010).

\bibitem{Drukier2012}
C. Drukier, L. Bartosch, A. Isidori, and P. Kopietz,
Phys. Rev. B {\bf 85}, 245120 (2012).

\bibitem{Holder2015}
T. Holder and W. Metzner,
Phys. Rev. B {\bf 92}, 041112 (2015).

\bibitem{Schattner2015}
Y. Schattner, S. Lederer, S. A. Kivelson, E. Berg,
Phys. Rev. X {\bf 6}, 031028 (2016).

\bibitem{cuprate-nfl}
se, e.g.,
A. Damascelli, Z.-X. Shen, and Z. Hussain,
Rev. Mod. Phys. {\bf 75}, 473 (2003);
D. Basov and T. Timusk, Rev. Mod. Phys. {\bf 77}, 721 (2005).


\bibitem{nfl-fesc}
see, e.g.,
Y. M. Dai, B. Xu, B. Shen, H. Xiao, H. H. Wen, X. G. Qiu, C. C. Homes, and R. P. S. M. Lobo,
Phys. Rev. Lett. {\bf 111}, 117001 (2013);
P. Palmsley \emph{et al.},  Phys. Rev. Lett. {\bf 110}, 257002 (2013).

\bibitem{Landau-Lifshitz}
see, e.g., L. D. Landau and E. M. Lifshitz,
\emph{Theory of Elasticity} (Pergamon Press, Oxford, 1970).

\bibitem{supplemental}
see Supplemental Material for technical details, which includes
Refs~\cite{Born1954,Chaplik1971}.

\bibitem{Born1954}
M. Born and K. Huang, \emph{Dynamics Theory of Crystal Lattices}
(Oxford University Press, Oxford, UK, 1954).

\bibitem{Chaplik1971}
A. V. Chaplik, Sov. Phys. JETP {\bf 33}, 997 (1971).

\bibitem{Cowley1976}
R. A. Cowley, Phys. Rev. B {\bf 13}, 4877 (1976).

\bibitem{Larkin1969}
A. I. Larkin and D. E. Khmelnitskii,
Sov. Phys. JETP {\bf 29}, 1123 (91969).

\bibitem{Levanyuk1970}
A. P. Levanyuk and A. A. Sobyanin, JETP Lett. {\bf 11}, 371 (1970).

\bibitem{Villain1970}
J. Villain, Solid State Comm. {\bf 8}, 295 (1970).

\bibitem{Cano2010}
A. Cano, M. Civelli, I. Eremin, and I. Paul,
Phys. Rev. B {\bf 82}, 020408(R) (2010).

\bibitem{Karahasanovic2016}
U. Karahasanovic and J. Schmalian,
Phys. Rev. B {\bf 93}, 064520 (2016).

\bibitem{Folk1976}
R. Folk, H. Iro, and F. Schwabl,
Z. Phys. B {\bf 25}, 69 (1976).

\bibitem{Zacharias2015}
M. Zacharias, I. Paul, and M. Garst,
Phys. Rev. Lett. {\bf 115}, 025703 (2015).

\bibitem{Zacharias2009}
M. Zacharias, P. W\"{o}lfle, and M. Garst,
Phys. Rev. B {\bf 80}, 165116 (2009).

\bibitem{gegenwart2006}
P. Gegenwart, F. Weickert, M. Garst, R. S. Perry, and Y. Maeno,
Phys. Rev. Lett. {\bf 96}, 136402 (2006).

\bibitem{rost2011}
A. W. Rost, S. A. Grigera, J. A. N. Bruin, R. S. Perry, D. Tian, S. Raghu, S. A. Kivelson, and A. P. Mackenzie,
 Proc. Natl. Acad. Sci. (USA) {\bf 108}, 16549 (2011).

\bibitem{gallais2013}
Y. Gallais, R. M. Fernandes, I. Paul, L. Chauvi\`{e}re, Y.-X. Yang, M.-A. M\'{e}asson, M. Cazayous, A. Sacuto, D. Colson,
and A. Forget,
Phys. Rev. Lett. {\bf 111}, 267001 (2013).

\bibitem{brouet2010}
V. Brouet, \emph{et al.}, Phys. Rev. Lett. {\bf 105}, 087001 (2010).

\end{thebibliography}
\end{document}